\documentclass[aps,prl,twocolumn,showpacs,superscriptaddress,nobibnotes,nofootinbib,showkeys,10pt]{revtex4-2} %,subfigure
\usepackage[dvips]{graphicx}
\usepackage{epsfig}
\usepackage{bm}   % for adding bold symbols etc. in maths mode
\usepackage{dcolumn}% Align table columns on decimal point
\usepackage{rotate}
\usepackage[table]{xcolor}
\usepackage{footmisc,booktabs,amssymb,longtable}
\usepackage{amsmath,amsfonts}

\usepackage{multirow}
\definecolor{Gray}{gray}{0.85}
\definecolor{LightCyan}{rgb}{0.88,1,1}

\usepackage[scaled=0.92]{helvet}
\usepackage[T1]{fontenc}
\usepackage{textcomp}
\usepackage{braket}
\usepackage{textgreek}
\usepackage{rotate}
\usepackage{tabularx}
\usepackage[referable]{threeparttablex}
\usepackage{hyperref}
\definecolor{capri}{rgb}{0.0, 0.75, 1.0}
\definecolor{cornflowerblue}{rgb}{0.39, 0.58, 0.93}
\definecolor{spirodiscoball}{rgb}{0.06, 0.75, 0.99}
\definecolor{pear}{rgb}{0.82, 0.89, 0.19}

\begin{document}

% \title{Nuclear Polarizability Behind Long-Standing Puzzle}
\title{Large quadrupole deformation in $^{20}$Ne challenges rotor model and modern theory:\\ urging for $\alpha$ clusters in nuclei}
%
%\title{Strong quadrupole collectivity  Coulomb-excitation measurement of the $B(E2;2^+_1 \rightarrow 0^+_1)$ value in  $^{20}$Ne} 
% `safe' measurement of the 2$^+_1$ quadrupole moment in $^{20}$Ne}

%Determination of $2^+_1 \rightarrow 0^+_1$ $E2$ transition strengths
%in $^{116}$Sn and $^{118}$Sn:\\

\author{C. V. Mehl}
\affiliation{Department of Physics \& Astronomy, University of the Western Cape, P/B X17, Bellville 7535, South Africa}

\author{J. N. Orce}
\thanks{Corresponding author: jnorce@uwc.ac.za}
\email{coulex@gmail.com} \homepage{http://nuclear.uwc.ac.za; https://github.com/UWCNuclear}
% \affiliation{TRIUMF, 4004 Wesbrook Mall, Vancouver, BC V6T 2A3, Canada}
\affiliation{Department of Physics \& Astronomy, University of the Western Cape, P/B X17, Bellville 7535, South Africa}
\affiliation{National Institute for Theoretical and Computational Sciences (NITheCS), South Africa}

\author{C. Ngwetsheni}
\affiliation{Department of Physics \& Astronomy, University of the Western Cape, P/B X17, Bellville 7535, South Africa}

\author{P. Marevi\'c}
\affiliation{Department of Physics, Faculty of Science, University of Zagreb, Bijeni\v{c}ka c. 32, 10000 Zagreb, Croatia}
% \affiliation{Nuclear and Chemical Sciences Division, Lawrence Livermore National Laboratory, Livermore, CA 94551, USA}
% \affiliation{CEA, DAM, DIF, F-91297 Arpajon, France}
% \affiliation{Institut de Physique Nucl\'eaire,
% Universit\'e Paris-Sud,
% IN2P3-CNRS, Universit\'e Paris-Saclay, F-91406 Orsay Cedex, France}

\author{B. A. Brown}
\affiliation{Department of Physics \& Astronomy and National Superconducting Cyclotron Laboratory, Michigan State University, East Lansing, MI 48824-1321, USA}

\author{J. D. Holt}
\affiliation{TRIUMF, 4004 Wesbrook Mall, Vancouver, BC V6T 2A3, Canada}
\affiliation{Department of Physics, McGill University, Montr\'eal, QC H3A 2T8, Canada}

\author{M. Kumar Raju}
\altaffiliation{Present address: Department of Physics, GITAM School of Science, GITAM University, Visakhapatnam 530045, India}
\affiliation{Department of Physics \& Astronomy, University of the Western Cape, P/B X17, Bellville 7535, South Africa}

\author{\\E. A. Lawrie}
\affiliation{iThemba LABS, National Research Foundation, P.O. Box 722, Somerset West 7129, South Africa}
\affiliation{Department of Physics \& Astronomy, University of the Western Cape, P/B X17, Bellville 7535, South Africa}

\author{K. J. Abrahams}
\affiliation{Department of Physics \& Astronomy, University of the Western Cape, P/B X17, Bellville 7535, South Africa}

\author{P. Adsley}
\affiliation{iThemba LABS, National Research Foundation, P.O. Box 722, Somerset West 7129, South Africa}
\affiliation{Department of Physics, Stellenbosch University, P/B X1, 7602 Matieland, South Africa}

\author{E. H. Akakpo}
\affiliation{Department of Physics \& Astronomy, University of the Western Cape, P/B X17, Bellville 7535, South Africa}

\author{R. A. Bark}
\affiliation{iThemba LABS, National Research Foundation, P.O. Box 722, Somerset West 7129, South Africa}

\author{N. Bernier}
\affiliation{Department of Physics \& Astronomy, University of the Western Cape, P/B X17, Bellville 7535, South Africa}
\affiliation{National Institute for Theoretical and Computational Sciences (NITheCS), South Africa}

\author{T. D. Bucher}
\affiliation{Department of Physics \& Astronomy, University of the Western Cape, P/B X17, Bellville 7535, South Africa}
\affiliation{iThemba LABS, National Research Foundation, P.O. Box 722, Somerset West 7129, South Africa}

\author{\\W. Yahia-Cherif}
\affiliation{Universit\'{e} des Sciences, Facult\'{e} de Physique, BP 32 El-Alia, 16111 Bab-Ezzouar, Alger, Algeria}

\author{T. S. Dinoko}
\altaffiliation{Present address: National Metrology Institute of South Africa, P/B X 34, Lynnwood Ridge, Pretoria 0040, South Africa.}
\affiliation{Department of Physics \& Astronomy, University of the Western Cape, P/B X17, Bellville 7535, South Africa}
\affiliation{iThemba LABS, National Research Foundation, P.O. Box 722, Somerset West 7129, South Africa}

\author{J.-P. Ebran}
\affiliation{CEA, DAM, DIF, F-91297 Arpajon, France}

\author{N. Erasmus}
\affiliation{Department of Physics \& Astronomy, University of the Western Cape, P/B X17, Bellville 7535, South Africa}

\author{P. M. Jones}
\affiliation{iThemba LABS, National Research Foundation, P.O. Box 722, Somerset West 7129, South Africa}

\author{E. Khan}
\affiliation{Institut de Physique Nucl\'eaire,
% Universit\'e Paris-Sud,
IN2P3-CNRS, Universit\'e Paris-Saclay, F-91406 Orsay Cedex, France}

\author{N. Y. Kheswa}
\affiliation{iThemba LABS, National Research Foundation, P.O. Box 722, Somerset West 7129, South Africa}
\affiliation{Department of Physics \& Astronomy, University of the Western Cape, P/B X17, Bellville 7535, South Africa}

\author{N. A. Khumalo}
\affiliation{Department of Physics \& Astronomy, University of the Western Cape, P/B X17, Bellville 7535, South Africa}
\affiliation{iThemba LABS, National Research Foundation, P.O. Box 722, Somerset West 7129, South Africa}

\author{J. J. Lawrie}
\affiliation{Department of Physics \& Astronomy, University of the Western Cape, P/B X17, Bellville 7535, South Africa}
\affiliation{iThemba LABS, National Research Foundation, P.O. Box 722, Somerset West 7129, South Africa}

\author{S. N. T. Majola}
\altaffiliation{Present address: Department of Physics, University of Johannesburg, P/B 524, Auckland Park, 2006, South Africa}
\affiliation{Department of Physics \& Engineering, University of Zululand, P/B X1001, KwaDlangezwa 3886, South Africa}

\author{K. L. Malatji}
\affiliation{Department of Physics \& Astronomy, University of the Western Cape, P/B X17, Bellville 7535, South Africa}
\affiliation{iThemba LABS, National Research Foundation, P.O. Box 722, Somerset West 7129, South Africa}

\author{D. L. Mavela}
\affiliation{Department of Physics \& Astronomy, University of the Western Cape, P/B X17, Bellville 7535, South Africa}

\author{M. J. Mokgolobotho}
\affiliation{Department of Physics \& Astronomy, University of the Western Cape, P/B X17, Bellville 7535, South Africa}

\author{\\T. Nik\v{s}i\'c}
\affiliation{Department of Physics, Faculty of Science, University of Zagreb, Bijeni\v{c}ka c. 32, 10000 Zagreb, Croatia}

\author{S. S. Ntshangase}
\affiliation{Department of Physics \& Engineering, University of Zululand, P/B X1001, KwaDlangezwa 3886, South Africa}

\author{V. Pesudo}
\affiliation{Department of Physics \& Astronomy, University of the Western Cape, P/B X17, Bellville 7535, South Africa}
\affiliation{iThemba LABS, National Research Foundation, P.O. Box 722, Somerset West 7129, South Africa}

\author{B. Rebeiro}
\affiliation{Department of Physics \& Astronomy, University of the Western Cape, P/B X17, Bellville 7535, South Africa}

\author{O. Shirinda}
\affiliation{iThemba LABS, National Research Foundation, P.O. Box 722, Somerset West 7129, South Africa}
\affiliation{Department of Physics, Stellenbosch University, P/B X1, 7602 Matieland, South Africa}

\author{D. Vretenar}
\affiliation{Department of Physics, Faculty of Science, University of Zagreb, Bijeni\v{c}ka c. 32, 10000 Zagreb, Croatia}

\author{M. Wiedeking}
\affiliation{iThemba LABS, National Research Foundation, P.O. Box 722, Somerset West 7129, South Africa}

%
%
% \address[a]{Department of Physics \& Astronomy, University of the Western Cape, P/B X17, Bellville ZA-7535, South Africa}
%
% \address[l]{National Institute for Theoretical and Computational Sciences (NITheCS), South Africa}
%
% \address[k]{Department of Physics, Faculty of Science, University of Zagreb, Bijeni\v{c}ka c. 32, 10000 Zagreb, Croatia}
%
% \address[b]{Nuclear and Chemical Sciences Division, Lawrence Livermore National Laboratory, Livermore, CA 94551, USA}
% \address[c]{CEA, DAM, DIF, F-91297 Arpajon, France}
% \address[d]{Institut de Physique Nucl\'eaire,
% % Universit\'e Paris-Sud,
% IN2P3-CNRS, Universit\'e Paris-Saclay, F-91406 Orsay Cedex, France}
%
% \address[e]{Department of Physics \& Astronomy and National Superconducting Cyclotron Laboratory, Michigan State University, East Lansing, MI 48824-1321, USA}
%
%
% \address[f]{TRIUMF, 4004 Wesbrook Mall, Vancouver, BC V6T 2A3, Canada}
%
% \address[i]{Department of Physics, McGill University, Montr\'eal, QC H3A 2T8, Canada}
%
%
% \address[g]{iThemba LABS, National Research Foundation, P.O. Box 722, Somerset West 7129, South Africa}
%
% \address[h]{Department of Physics, Stellenbosch University, P/B X1, 7602 Matieland, South Africa}
%
%
%
%
% \address[j]{Department of Physics \& Engineering, University of Zululand, P/B X1001, KwaDlangezwa 3886, South Africa}
%
% \address[m]{Universit\'{e} des Sciences, Facult\'{e} de Physique, BP 32 El-Alia, 16111 Bab-Ezzouar, Alger, Algeria}
%

\date{\today}

\begin{abstract}

% \noindent

The  spectroscopic quadrupole moment of the first excited state, $Q_{_S}(2^{+}_{1})$, at 1.634 MeV in $^{20}$Ne
was determined from sensitive reorientation-effect Coulomb-excitation measurements using a heavy target and safe energies well below the Coulomb barrier.
Particle-$\gamma$ coincidence measurements were collected at iThemba LABS with a digital data-acquisition system using the {\sc AFRODITE} array coupled to an annular, doubled-sided silicon detector.
A precise value of $Q_{_S}(2^{+}_{1})=-0.22(2)$ eb was determined at backward angles in agreement with the only safe-energy measurement prior to
this work, $Q_{_S}(2^{+}_{1})=-0.23(8)$ eb.
This result adopts 1$\hbar\omega$ shell-model calculations of the nuclear dipole polarizability of the 2$^+_1$
state that contributes to the effective quadrupole interaction and determination of $Q_{_S}(2^{+}_{1})$.
It disagrees, however, with the ideal rotor model for axially-symmetric nuclei by almost $3\sigma$.
% , one of the largest discrepancies found in nuclei.
% The discrepancy with macroscopic and microscopic models is considerable with the ideal rotor model underestimating the quadrupole moment by $\approx 1.65\sigma$.
Larger discrepancies are computed by modern state-of-the-art calculations performed in this and prior work,
including {\it ab initio} shell model with chiral effective interactions and the multi-reference relativistic energy density functional ({\sc MR-EDF}) model. The intrinsic nucleon density of the 2$^+_1$ state in $^{20}$Ne calculated with the {\sc MR-EDF} model illustrates the presence of $\alpha$ clustering, which explains the largest discrepancy with the rotor model found
in the nuclear chart
% and/or related triaxiality effects
and motivates the explicit  inclusion of $\alpha$ clustering for full convergence of $E2$ collective properties.

\end{abstract}

\keywords{spectroscopic quadrupole moment, nuclear dipole polarizability, safe Coulomb excitation, B(E2) values, photo-absorption cross sections, {\it ab initio}, mean-field and shell models, $\alpha$ clusters}

\maketitle

Light-ion collisions ({\sc LIC}) is a novel avenue of research at modern heavy-ion collision ({\sc HIC}) facilities~\cite{lic}
aimed at understanding the general hydrodynamic behaviour of the quark-gluon plasma and the geometry of light nuclei,
where varying shapes --- e.g., $\alpha$ clusters and triaxiality fluctuations --- may be inferred
from the  overlap region or centrality~\cite{aaij2022centrality} of the two colliding nuclei~\cite{brewer2021opportunities,giacalone2024unexpected}.
The application to smaller symmetric systems --- e.g., $^{16}$O+$^{16}$O and $^{20}$Ne+$^{20}$Ne --- has an advantage with respect to
typical  Pb+Pb, Au+Au and other {\sc HIC}s involving protons because of the less sensitivity to spatial fluctuations
in the proton and
% less eccentric geometry, i.e.,
a better defined centrality~\cite{shen2751277dynamical}.
One of the most crucial shapes concerns the nucleus $^{20}$Ne~\cite{ebran2012atomic},
where a cluster formation --- the so-called bowling pin --- is expected from an increased incoherent cross section
relative to an assumed spherical shape~\cite{mantysaari2023multiscale}. The initial geometry of the nucleus before collision is crucial, nonetheless, to assess nuclear clusters in future {\sc LIC} measurements~\cite{giacalone2024unexpected}.

A long-standing challenge to nuclear structure physics concerns the large quadrupole deformation
of the first excited state,  $J^{\pi}=2^+_1$, in $^{20}$Ne~\cite{thompson} --- where $J$ is the angular momentum and $\pi$ the parity ---
% , with angular momentum $J^{\pi}=2^+_1$, 
% at 1.633 MeV  
% one of the earliest isotopes being discovered:
% Particularly,
% Fig.~\ref{fig:theoryvsexp2} suggests that 
which arguably presents the most extreme quadrupole shape in the Segr\'e chart.
% of any even-even nucleus in the $sd$-shell,
% between proton and neutron magic numbers 8 and 20.
With only two protons and two neutrons above the shell closure with magic number 8, 
a favorable prolate deformation at the beginning of the $sd$ shell is expected from the relatively 
sharper decrease in energy of the $\frac{1}{2}[220]$ single-particle Nilsson orbit as a function of deformation~\cite{bm}. 
However, the magnitude of the quadrupole deformation cannot be reproduced by the rotational model of Bohr and Mottelson~\cite{bm} nor can it be computed by the shell~\cite{Spear,brown}, mean-field~\cite{20Ne_gogny,zhou,marevic2}, resonating group method~\cite{matsuse} and multi-configurational~\cite{lebloas} models. It further questions the simple picture provided by the $N_pN_n$ scheme~\cite{hamamoto}. With $N_p(N_n)$ being the number of valence proton (neutron) pairs,
a larger number of valence nucleons in midshell  should produce larger deformations within a major shell.

In more detail, the spectroscopic quadrupole moment in the laboratory frame, $Q_{_S}(J)$,
can be determined for a nuclear state with $J\neq0,\frac{1}{2}$ using the reorientation effect ({\sc RE})
arising from the hyperfine interaction between $Q_{_S}$ and the
time-dependent electric-field gradient generated by the projectile ({\sc P}) and target ({\sc T}) during the scattering process~\cite{hausser0,deb1,orce_review}.
Assuming an axially-symmetric rotor,  
the intrinsic quadrupole moment of a nucleus in the body-fixed frame, $Q_{_0}$, can be determined from  
the  reduced transition probability connecting the 0$^+_1$ ground and 2$^+_1$ states with an electric quadrupole ({\sc E2}) operator, or $B(E2; 0^+_1\rightarrow 2^+_1)$ value~\cite{bm}, as
\begin{equation}
Q_{_0}=\left(\frac{16\pi}{5}~B(E2; 0^+_1\rightarrow 2^+_1)\right)^{1/2}. 
\label{eq:rotmodel}
\end{equation}
% For the ground-state band with fixed quadrupole deformation $\beta$ and no triaxility, $\gamma=0^{\circ}$, 
% $Q_{_0}$ is related to $\beta$ as
% \begin{equation}
%   Q_{_0}=\left(\frac{16\pi}{5}\right)^{1/2} \frac{3}{4\pi} ~Ze R^2 \beta, 
%   \label{eq:beta}
% \end{equation}
% where $Z$ is the proton number and $R$, the nuclear radius, $R=1.2~ A^{1/3}$ fm and 
% $\beta\approx1.06~\delta\approx 0.945\frac{\Delta R}{R}$, with $\Delta R$ being the difference between the radii parallel and 
% perpendicular to the symmetry axis~\cite{bm}. 
% $B(E2; 0^+_1\rightarrow 2^+_1)$ value~\cite{nndc}. 
% For  arbitrary $J$ and $K$, the \emph{spectroscopic quadrupole moment} in the laboratory frame, 
% $Q_{_S}$, and $Q_{_0}$ can be related by 
% % \begin{equation}
% $Q_{_S}=\frac{3K^2-J(J+1)}{(2J+3)(J+1)}~Q_{_0}$,  
% % \label{eq:qsq0}
% %  \end{equation}
% which 
For  $J^{\pi}=2^+_1$ and $K=0$ (e.g., ground-state band), where $K$ is the projection of $J$ onto the symmetry axis,  $Q_{_S}(2^+_1)$ and $Q_{_0}$ can be related by
% $Q_{_S}(2^+_1)=-2/7~Q_{_0}$.
\begin{equation}
Q_{_S}(2^+_1)=\frac{3K^2-J(J+1)}{(2J+3)(J+1)}~Q_{_0}=-2/7~Q_{_0},
\label{eq:qsq0}
 \end{equation}
where $Q_{_S}(2^+_1)<0$ for prolate and $Q_{_S}(2^+_1)>0$ for oblate deformations.
% $Q_{_S}(2^+_1)=-2/7~Q_{_0}$. 
% which generates a time-dependent hyperfine splitting of the nuclear levels~\cite{hausser0,deb1}. 
% which enhances ({\sc $Q_{_S}>0$}) or inhibits ({\sc $Q_{_S}<0$})
% the Coulomb-excitation cross section~\cite{hausser0,deb1}. 
% hence providing a spectroscopic probe for a measurement of {\sc $Q_{_S}$}.
% at 1.634 MeV

Nuclear interference may substantially affect the measurement of $Q_{_S}(J)$ values,
a common scenario in most of the first {\sc RE} measurements carried out in the 1960s and 1970s~\cite{Spear}.
The nucleus $^{24}$Mg provides
one of the most dramatic cases~\cite{Spear}, where a very large $Q_{_S}(2^{+}_{1})=-24.3\pm 3.5$ eb was determined~\cite{hausser1969measurement},
and soon disregarded by some of the same authors~\cite{otto2} and other groups~\cite{fewell1979static} because of the very close distance between the colliding nuclei.
For negligible nuclear contributions, Spear's systematic study of $Q_{_S}(2^{+}_{1})$ values in $sd$-shell nuclei~\cite{Spear}
suggests a minimum safe distance of closest approach between nuclear surfaces of $S(\vartheta)_{min} \geq  6.5$ fm,
where
% $S(\vartheta)$ is defined as
\begin{eqnarray}
S(\vartheta)&=&\frac{e^2 Z_{_P} Z_{_T}}{8 \pi \epsilon_0 T_{lab}}(1+A_{_P}/A_{_T})\left[1+cosec(\vartheta/2)\right] \nonumber \\ &-&1.25(A_{_P}^{1/3}+A_{_T}^{1/3})~\mbox{fm}. 
\end{eqnarray}
Here, $\vartheta$ is the scattering angle in the center-of-mass frame, $Z$ and $A$ the proton and mass number, respectively,
and $T_{lab}$  the kinetic or beam energy in the laboratory frame.
Prior to this work,  three {\sc RE} Coulomb-excitation  measurements were carried out in $^{20}$Ne~\cite{Nakai,Schwalm,Olsen}.
% , with the last one dating from 1974~\cite{Olsen}.
% Such a paucity arises because of the difficulty associated with the production of neon ion beams in tandem accelerators.
Two of the previous {\sc RE} measurements,  yielding $Q_{_S}(2^{+}_{1})=-0.24(3)$ eb~\cite{Nakai}  and
$Q_{_S}(2^{+}_{1})=-0.20(5)$ eb~\cite{Olsen},  were carried out at exceedingly high $T_{lab}$ energies, with
 $S(\vartheta)_{min}$ as low as 3.8 and 4.2 fm, respectively.
% % and  failed to show that these Coulomb-excitation measurements 
% % could be deemed as safe. 
% % and consistent results~\cite{Spear}. 
The third {\sc RE} measurement by Schwalm and collaborators was performed using a $^{20}$Ne gas target at a safe  $S(\vartheta)_{min}=7.1$ fm, yielding $Q_{_S}(2^{+}_{1})=-0.23(8)$ eb.
% This rather large uncertainty is associated with the use of a $^{20}$Ne gas target~\cite{Schwalm} because of
% the sensitivity of the {\sc RE} is directly proportional to  $Z_{_T}/Z_{_{P}}$~\cite{deb1,Nakai}.
These  {\sc RE} measurements give a weighted average of  $Q_{_S}(2^+_1)=-0.23(3)$ eb,
% for $^{20}$Ne~\cite{Schwalm,Nakai,Olsen} -- 
which is the  accepted value in the National Nuclear Data Center (NNDC)~\cite{Spear}. 
% The  $Q_{_S}(2^+_1)_{RE}=-0.23(3)$ eb value for the  2$^+_1$ state at 1.634 MeV in $^{20}$Ne yields 
% and   $Q_{_0}=+0.81(11)$ eb. 
Because of the relatively small  $Z$ and radius $R$, 
% Eq. \ref{eq:beta} 
this represents the largest quadrupole (prolate) deformations in the nuclear chart with a quadrupole parameter of
$\beta_2=0.96(11)$~\cite{bm}.

It is interesting to compare the $Q_{_S}(2^+_1)$ value determined with the {\sc RE}  and the one extracted from the rotational model~\cite{bm}, $Q_{_S}(2^+_1)_{B(E2)}$, using Eqs. \ref{eq:rotmodel} and \ref{eq:qsq0},
by defining the spectroscopic quadrupole ratio~\cite{robinson2006shell,yeager2008relations,rq},
\begin{equation}
r_q:=\left|\frac{Q_{_S}(2^+_1)}{Q_{_S}(2^+_1)_{B(E2)}}\right|.
\label{eq`:rq}
\end{equation}
% where  $Q_{_S}(2^+_1)_{B(E2)}$ is determined using from the rotational model 
% (Eqs. \ref{eq:rotmodel} and \ref{eq:qsq0})~\cite{bm}. 
% (Eqs.~\ref{eq:rotmodel} and \ref{eq:qsq0})~\cite{bm}. 
Data show $r_q\approx 1$ for good rotors in the rare-earth and $A\sim 180$ regions,
while $r_q= 0$ is expected for an ideal vibrator ($Q_{_S}(2^+_1)=0$)~\cite{bm}.
With $Q_{_S}(2^+_1)_{B(E2)} = \pm 0.165(4)$ eb~\cite{raman,schwalm3,20ne_kenn,Spear},
an anomalously large $r_q(2^+_1)=1.39(49)$ is determined in $^{20}$Ne considering the only safe {\sc RE} measurement.
% despite the fact that $E(4^+_1)/E(2^+_1)=2.6$ is not typical of rotor or vibrators. 
% making this one of the few known cases
% in the nuclear chart 
% where $Q_{_S}(2^+_1)$ and $Q_{_S}(2^+_1)_{B(E2)}$ values disagree at the $2.15\sigma$ confidence level.
As already stated in Spear's 1981 review article~\cite{Spear}, a new measurement of the $Q_{_S}(2^+_1)$ value in $^{20}$Ne
was desirable.
We present it here from a safe, precise and sensitive {\sc RE} measurement.
% that quantifies the competing contribution from the nuclear electric dipole polarizability.

% \begin{figure}[]
% \begin{center}
% \includegraphics[width=7.cm,height=4.5cm,angle=-0]{Tigress2.eps}
% \caption{(Color online) TIGRESS $\gamma$-ray array comprising 
% eight HPGe clover detectors and full BGO Comptom-suppression shields mounted to either side of the 
% {\normalsize B{\footnotesize AMBINO}} target chamber at the TRIUMF facility. A different configuration of 
% six HPGe clover detectors was employed for the $^{20}$Ne experiment described in this work.} 
% \label{fig:1}
% \end{center}
% \end{figure}

% 
% Modern high-efficient and segmented HPGe $\gamma$-ray arrays such as TIGRESS~\cite{tigress} at the TRIUMF facility 
% are able to perform Coulomb-excitation studies of incident light projectiles with high-lying 2$^+_1$ states. 
%For consistency, $Q_s(2^+_1)=-0.23(8)$ eb should rather be the accepted value in the NNDC database 
%and previous discrepancies would dissapear~\cite{nndc}.  
% 
% Such an uncertain scenario does not permit to draw any conclusions and a more accurate measurement 
% of $Q_{_S}(2^+_1)$ in $^{20}$Ne is desirable. 

% The safe Coulomb excitation of $^{20}$Ne$^{3+}$ beams at 73 MeV used to obtain the measurement of the $\langle2^{+}_{1}||E2||2^{+}_{1}\rangle$ diagonal 
% matrix element in $^{20}$Ne was carried out 
% at iThemba LABS over the course of three days. 

Coulomb-excitation measurements have been carried out at iThemba {\sc LABS}
using the $^{194}$Pt($^{20}$Ne,$^{20}$Ne$^*$)$^{194}$Pt$^*$ reaction at a safe energy of 71.3 MeV
and a digital {\sc DAQ} system based on 100 MHz Pixie-16 modules from {\sc XIA LLC}~\cite{xia}.
% over the course of three days. 
Pure beams of $^{20}$Ne$^{4+}$ ions  at $\approx1\times10^9$ pps were extracted from a 99.99\%-enriched $^{20}$Ne gas bottle and accelerated with the $K=200$ separated sector cyclotron (SSC) onto a 95$\%$-enriched $^{194}$Pt target of 1.2 mg/cm$^{2}$ thickness.
The scattered  ions were detected using an annular, double-sided silicon detector ({\sc S3}-type from Micron Semiconductors~\cite{cat}) comprising  24 rings and 32 sectors~\cite{Ostrowski} and
mounted upstream at 27.5 mm from the target position and perpendicularly aligned with the beam axis.
The scattering $\theta$ angles in the laboratory frame ranged between 128.1$^{\circ}$ and  157.3$^{\circ}$,
corresponding to $S(\vartheta)\approx 7.5$ and 6.9 fm, respectively. 
Gamma rays were collected with the {\sc AFRODITE} array~\cite{afrodite} comprising eight HPGe clover detectors
at a distance of 19.6 cm between target and Ge crystals.
% emitted 
% in the de-excitation of states 
% in $^{20}$Ne and $^{194}$Pt, 
% where the faces of the clover detectors subtended ($\theta_{\gamma}, \phi_{\gamma}$)
% laboratory angles of (90$^{\circ}$, 45$^{\circ}$), (90$^{\circ}$, 225$^{\circ}$), (90$^{\circ}$, 270$^{\circ}$), (90$^{\circ}$, 315$^{\circ}$), (135$^{\circ}$, 0$^{\circ}$), (135$^{\circ}$, 90$^{\circ}$), (135$^{\circ}$, 180$^{\circ}$) and (135$^{\circ}$, 270$^{\circ}$),
% in a right-handed coordinate system with the $z$-axis downstream of the beam direction.

% \begin{figure}[ht]
% \begin{center}
% \vspace{0.3cm}
% \includegraphics[width=6.7cm,height=4cm,angle=-0]{ring1_24gates.eps}\\
% \vspace{0.1cm}
% \includegraphics[width=6.7cm,height=4.3cm,angle=-0]{ringgammatimediff.eps}
% \caption{(Top panel) Particle spectra for the innermost [128.1$^{\circ}$, 128.8$^{\circ}$] (orange) and outermost
% [155.8$^{\circ}$, 157.3$^{\circ}$] (black) rings.
% % The elastic peaks correspond to 63.8 MeV and 60.9 MeV, respectively.
% A particle gate marked by the  dashed line was used to separate
% $^{20}$Ne from $^{194}$Pt ions. (Bottom panel) Time spectra (10 ns/channel) showing the time difference between particle hits in the rings and
% $\gamma$ rays as well as the gates taken for the prompt (solid lines) and background subtraction (dashed lines).}
% % Prompt and background gates of $\approx$60 ns were utilized to suppress random events.}
% \label{fig:1}
% \end{center}
% \end{figure}

\begin{figure}[]
\begin{center}
\vspace{0.2cm}
\includegraphics[width=7.5cm,height=5cm,angle=-0]{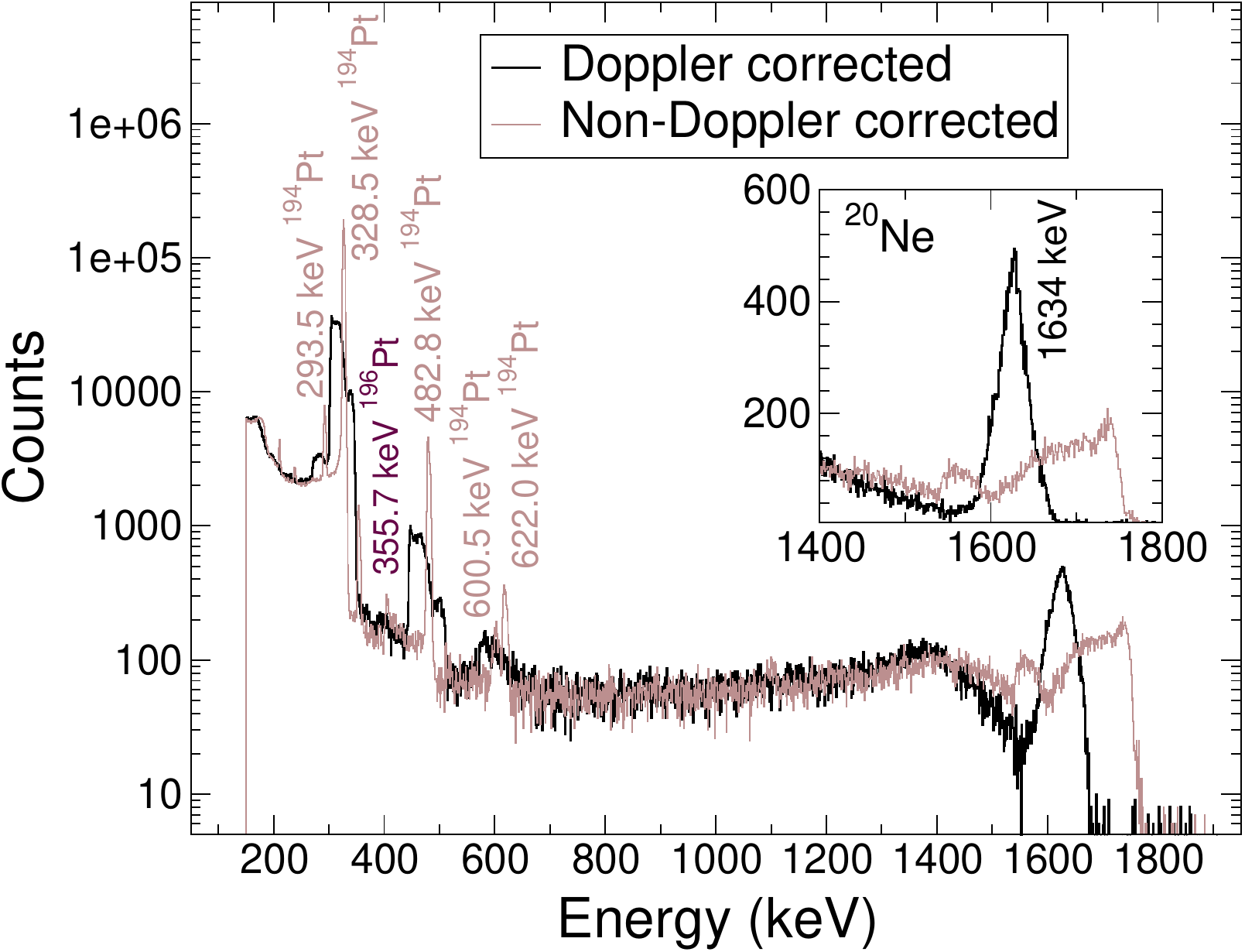}
\caption{Doppler (black) and non-Doppler (brown) corrected $\gamma$-ray energy spectra in log-y scale for the
$^{194}$Pt($^{20}$Ne,$^{20}$Ne$^*$)$^{194}$Pt$^*$ reaction at 71.3 MeV. The inset shows the
1634-keV peak in $^{20}$Ne in a linear scale.}
% at 71.3 MeV.}
% Spectra arise as the result of particle-$\gamma$ coincidences, Doppler and non-Doppler correction,
% energy-sharing conditions, add-back and particle and time gates.}
\label{fig:2}
\end{center}
\end{figure}

% \vspace{-1cm}
A low background $\gamma$-ray spectrum is achieved by collecting particle-$\gamma$ coincidence events, which require
% % This coincidence requirement was satisfied by implementing:
% % various conditions in a fast sorting code that has been developed for Coulomb excitation  data analysis at iThemba LABS, 
% % which include: 
% 1)  a broad particle energy gate covering the range of the scattered $^{20}$Ne$^{4+}$ ions, 
% % , as  shown in the top left panel of Fig.~\ref{fig:2}, 
% % This condition was used to reduce background in the particle-$\gamma$ energy spectra. 
% % In this case, the particle energy gate ranged from 48 to 81 MeV as shown in the top panel of Fig.~\ref{fig:1}, which shows the broad energy gate (green) scaled
% % by a factor of 10, covering the range of the elastic peak energies of ring 1 (shown in blue) and
% % ring 24 (shown in orange). 
% 2) particle-$\gamma$ time gates 
% % in the time difference spectrum, 
% % as shown in the bottom left panel of Fig.~\ref{fig:2}. 
% where a coincidence between 
both a $\gamma$ and a particle hit within a coincidence time window of approximately 600 ns.
% (10ns/channel).
% , as shown in the bottom panel of Fig.~\ref{fig:1}.
A gate outside this prompt
% particle-$\gamma$
window was used for background subtraction.
% of the spectrum produced by gating on the prompt particle-$\gamma$ window.
The particle hit was defined by a signal in one $\theta$ ring and one $\phi$ sector.
A broad particle energy gate covering the range of the scattered $^{20}$Ne$^{4+}$ ions was also implemented.
% as shown in the top panel of Fig.~\ref{fig:1}.
% Events outside this window are considered to be random coincidences and are 
% background subtracted.
% (background gate in Fig.~\ref{fig:1}).
% 3) charge sharing arising from incomplete charge collection between
% % as the full energy of a particle may be shared between 
% rings and sectors active and dead layers. This was achieved by  requiring that $|E_{_{sector}}$ - $E_{_{ring}}|<500$ keV,  
% % to be less than a energy chosen to reduce the background,
% while conserving the counts in the $\gamma$-ray peaks of interest.
% and 4)  additional inelastic gates on the rings. 
% The $\theta_{p}$ and $\phi_{p}$ angles as well as the $\theta_{\gamma}$ and $\phi_{\gamma}$ angles at the center of the crystal in 
% which a valid $\gamma$-ray  event occurred 
% were then used
% to calculate the relative angle between a $\gamma$ ray and a particle. 
When multiple  $\gamma$-ray events were detected in crystals within the same clover, the crystal with the highest deposited
energy was used to Doppler correct the add-back energy of the detected $\gamma$ ray. 
Figure~\ref{fig:2} shows the resulting Doppler and non-Doppler corrected $\gamma$-ray
energy spectra in coincidence with all rings. Total counts of $N_{_\gamma}^P=19523(953)$
% 45776(231)$
and $N_{_\gamma}^T=882606(29334)$,
% 493278(696)$
% for the 1634- and 328-keV $\gamma-$ray transitions,
respectively,  for the  1634- and 328-keV  $\gamma$ rays
depopulating the 2$^+_1$ states in $^{20}$Ne and $^{194}$Pt were measured.

\begin{figure}[!ht]
\begin{center}
\includegraphics[width=8.4cm,height=5.2cm,angle=-0]{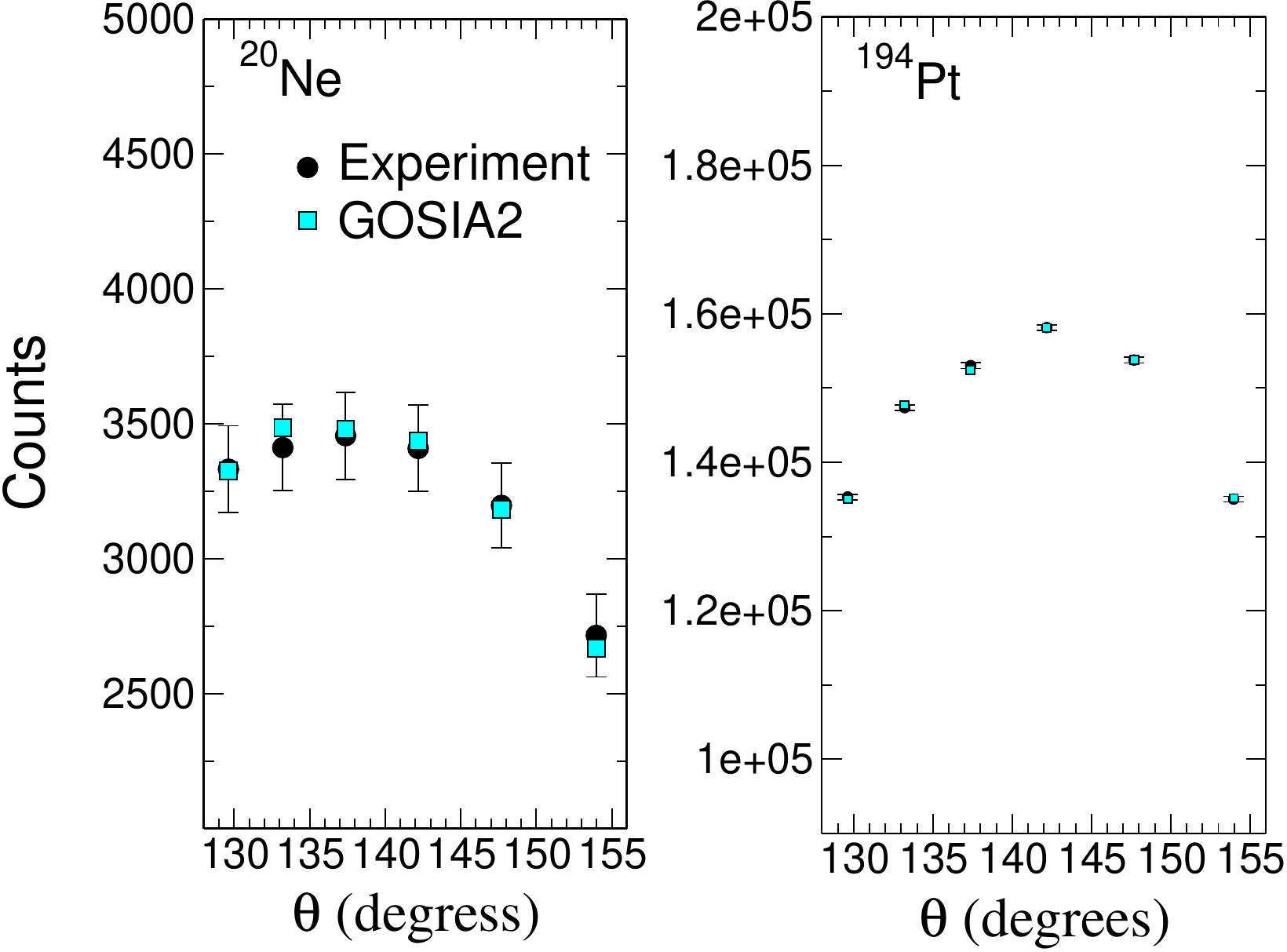}
\caption{Experimental and calculated ({\sc GOSIA}) $\gamma$-ray yields as a function of particle angle $\theta$ for the de-excitation of the 2$_1^+$ states in $^{20}$Ne (left) and $^{194}$Pt (right).}
\label{fig:3}
\end{center}
\end{figure}

The Coulomb-excitation analysis was performed using the semi-classical coupled-channel  least-squares codes
{\sc GOSIA} and {\sc GOSIA}2~\cite{gosia,magda}.
The integration of $\gamma$-ray yields accounts for detector angular limits, beam energy losses~\cite{srim} and the solid angle subtended by the silicon detector. The effect of higher-lying states in $^{20}$Ne was estimated using {\sc GOSIA} and considered negligible (<0.1\%).
% (with axial symmetry) 
% and the energy loss of the beam through the target.  
Integrated yields for each data set are constrained by available spectroscopic information concerning level lifetimes, 
branching ratios and matrix elements for all significant couplings up 
to the 3$^-_1$ states in $^{20}$Ne~\cite{nndc} and $^{194}$Pt~\cite{wu1996quadrupole,nndc}.
Further corrections include $\gamma$-ray efficiencies with  standard $^{152}$Eu and $^{56}$Co calibration sources
($\varepsilon_{_\gamma}^P = 145.3(4.4)$ and $\varepsilon_{_\gamma}^T = 344.5(6.9)$), internal-conversion processes~\cite{bricc}
and angular-distribution attenuation factors for the finite size of the detectors~\cite{gosia}.  
As shown in Fig.~\ref{fig:3}, experimental $\gamma$-ray yields per 4 rings were rebinned to obtain 6 data points and compared with integrated yields calculated with {\sc GOSIA},
which were normalized independently to the experimental yields in the projectile and target.
The shape of the angular distributions predicted by {\sc GOSIA} for both $^{20}$Ne and $^{194}$Pt
are in agreement with experiment,  and follow similar trends to those predicted by $d\sigma_{_{E2}}=P~d\sigma_{_R}$,
where $P$ is the probability of exciting the 2$^+_1$ state following the scattering process 
and $\sigma_{_{E2}}$ and $\sigma_{_R}$ are the Coulomb-excitation and Rutherford cross sections, respectively.
% Total experimental yields of 45776(231) and 493278(696) counts are observed for the 1634-  and 328-keV $\gamma$-ray peaks in  $^{20}$Ne and $^{194}$Pt, respectively. 

\begin{figure*}[!ht]
\begin{center}
\includegraphics[width=5cm,height=4.5cm,angle=-0]{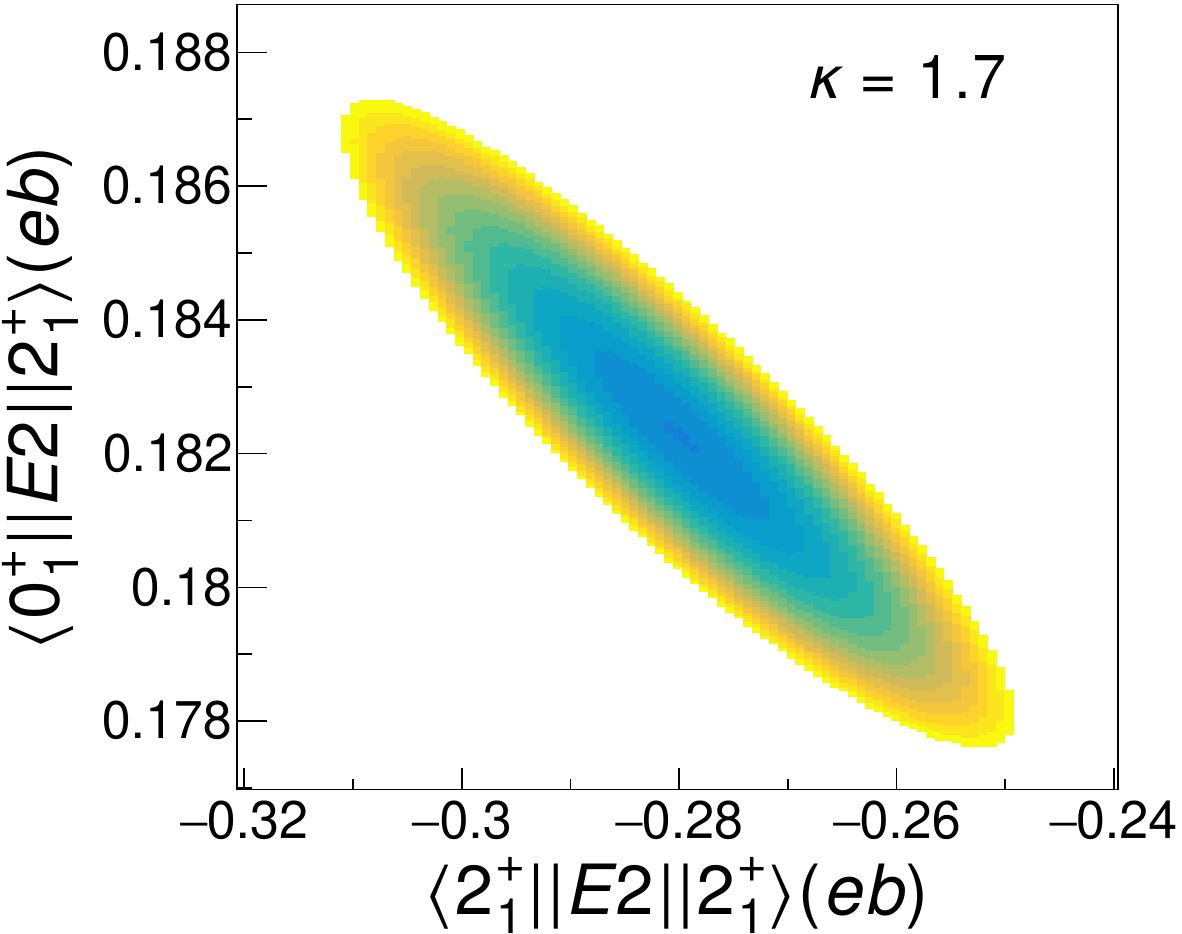}
\hspace{0.2cm}
\includegraphics[width=4.8cm,height=4.5cm,angle=-0]{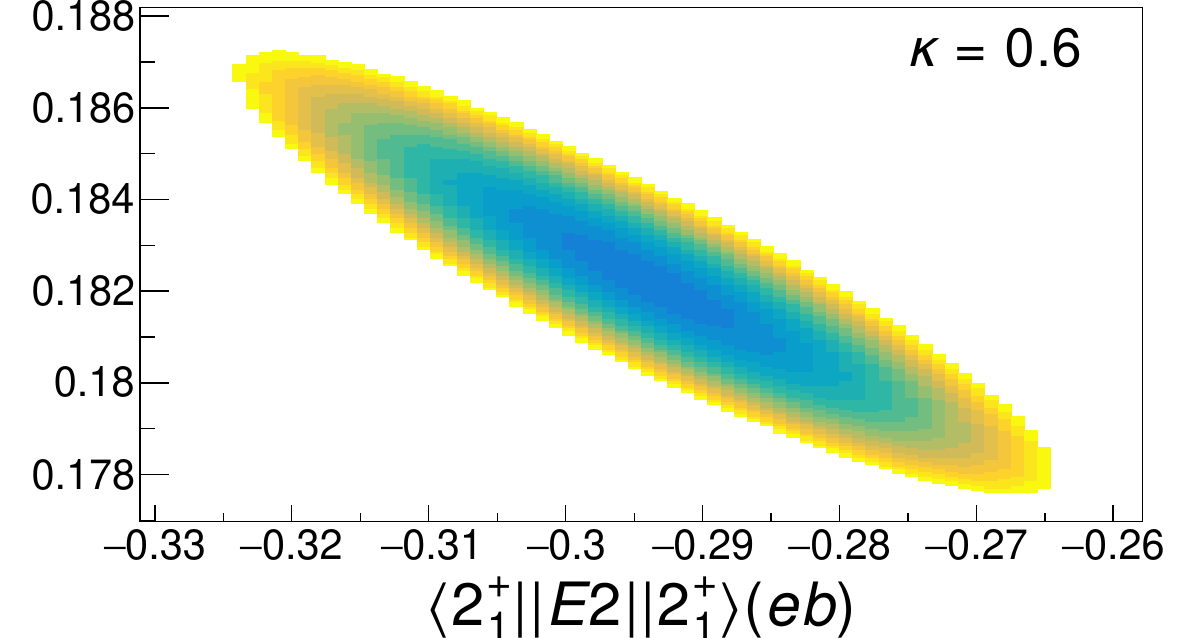}
\hspace{0.2cm}
\includegraphics[width=6cm,height=4.5cm,angle=-0]{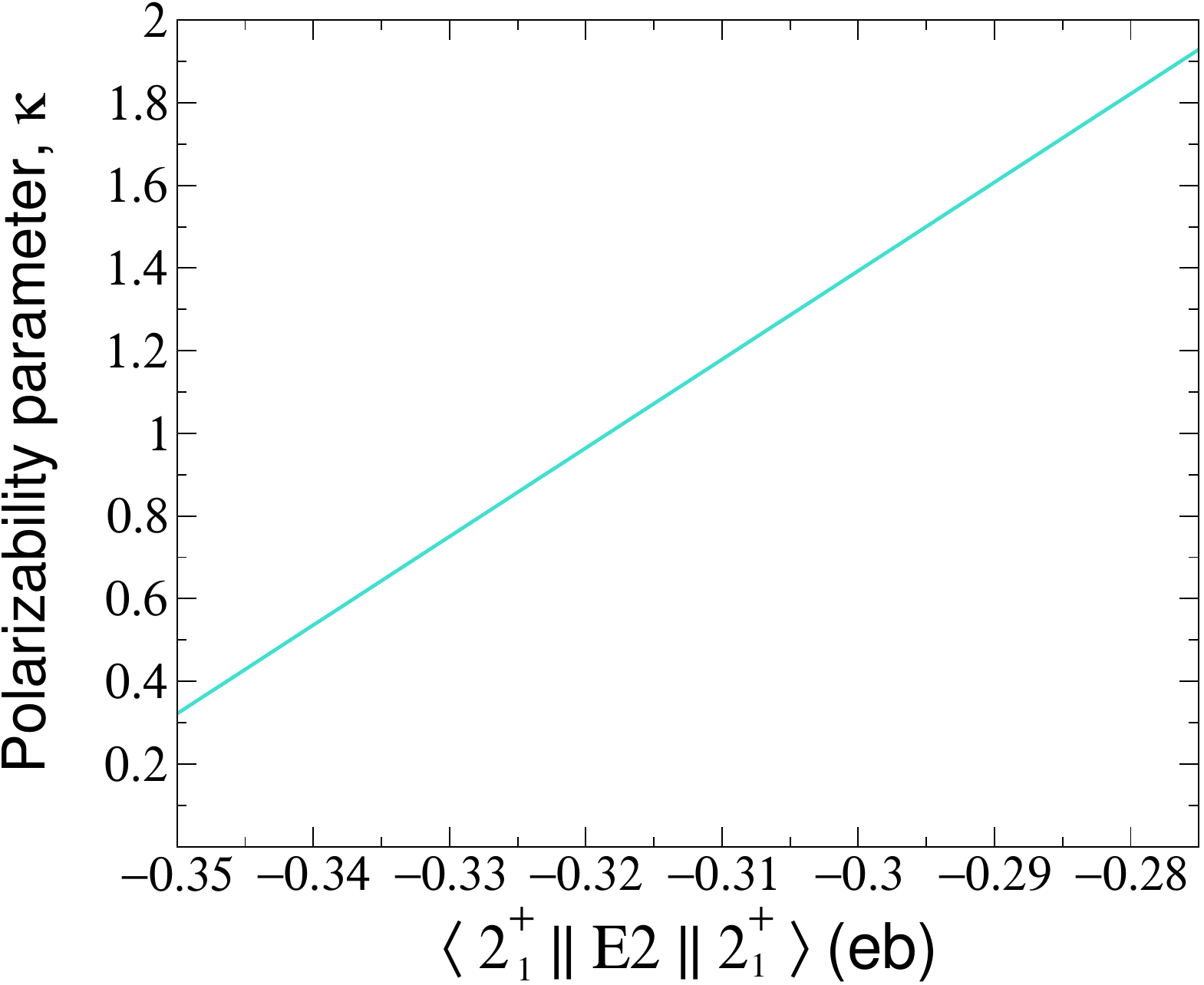}
% \includegraphics[width=7cm,height=3.cm,angle=-0]{20Ne_250x250}
% \vspace{-1cm}
% \includegraphics[width=0.6cm,height=2.8cm,angle=-0]{LongRedProlateNuc.eps}
\caption{Minimization of the $\chi^2$ surface in the  $\langle 2^+_1\mid\mid \hat{E2} \mid\mid 2^+_1\rangle$-$\langle 0^+_1\mid\mid \hat{E2} \mid\mid 2^+_1\rangle$ 2D-plane for the $^{20}$Ne projectile when combined with lifetime measurements~\cite{be2} for $k(2^+_1)=1.7$ (left) and $k(2^+_1)=0.6$ (right). The data are normalised to the excitation of the $^{194}$Pt target at a beam energy of 71.3 MeV using {\sc GOSIA}2.
}
\label{fig:chqsurface}
\end{center}
\end{figure*}

In order to determine the values of $\langle 2^+_1\mid\mid \hat{E2} \mid\mid 2^+_1\rangle$ and $\langle 2^+_1\mid\mid \hat{E2} \mid\mid 0^+_1\rangle$
that reproduce the experimental yields, the maximum likelihood approach~\cite{ceder} has been used.
Figure \ref{fig:chqsurface} shows the $\chi^{2}$ surface  with 300 $\times$ 300 scanned points together with the  corresponding $\chi^2 < \chi^2_{min} +1$ for polarizability parameters of $k(2^+_1)=1.7$ (left) and $k(2^+_1)=0.6$ (middle), as discussed below.
The data are normalised to the excitation of the $^{194}$Pt target at a beam energy of 71.3 MeV using GOSIA2~\cite{magda},
in conjunction with the accepted value of $\langle 2^+_1\mid\mid \hat{E2} \mid\mid 0^+_1\rangle=0.1825(44)$ eb in $^{20}$Ne~\cite{be2}, which is in agreement with the transitional matrix element extracted from a recent high-precision lifetime measurement~\cite{petkov}.

Further care must be taken in second-order Coulomb-excitation studies of light nuclei~\cite{eichler,deb1,hausser0,alder,orce_10Be}
where the  nuclear electric  dipole ({\sc E1})  polarizability  competes with the {\sc RE} and
may influence the determination  of $B(E2)$ and $Q_{_S}(2^+_1)$ values~\cite{eichler,orce_review}.
% \vspace{-1cm}. 
The {\sc E1} polarizability is directly related to the static polarizability, $\alpha_{_{E1}}=\frac{\mathbf{P}}{\mathbf{E}}$,
where {\bf P} is the electric dipole moment  induced in a nucleus by the time-dependent electric field {\bf E} of the partner.
Consequently, {\sc E1} virtual excitations via high-lying states in the isovector
giant dipole resonance ({\sc GDR})~\cite{berman1975} --- a collective motion of protons and neutrons out of phase
characterized by the residual
particle-hole interaction~\cite{brown1959dipole} ---
can polarize
the ground and excited states of nuclei~\cite{eichler,alder,levinger2} without involving short-range Yukawa forces. Particularly, two-step processes of the type $0^+_1 \rightarrow 1^-_{_{GDR}} \rightarrow 2^+_1$
% may polarize
% virtual electric-dipole excitations of 
can polarize the shape of the $2^+_1$ state.
% The  $E1$  polarizability arises from large $E1$ matrix elements 
% via virtual excitations of the isovector giant dipole resonance (\emph{\sc GDR})~\cite{berman1975}, i.e.
% For projectile excitation, the polarization potential $V_{pol}(t)$ is defined as~\cite{alder}
% \begin{equation}
%  V_{pol}(t)=0.0038 \kappa ~\frac{T_p A_p}{Z_p^2(1+A_p/A_t)}E_{(1,1,2)}(\vartheta,\xi),
%  \label{ref:finalpol}
% \end{equation}
% where the dipole polarization function, $E_{(1,1,2)}(\vartheta,\xi)$, is expressed in terms of orbital integrals~\cite{alder},
% and the polarizability parameter $\kappa$ is introduced to account for deviations of the actual {\sc GDR} effects
% % (circles in Fig.~\ref{fig:sigmasym})
% to those predicted by the hydrodynamic model~\cite{migdal1944quadrupole,levinger}.
% In our previous analysis, a value of $\kappa=1$ (orange band in Fig.~\ref{coulex}) was used in
% accord with previous work.
This effect is characterized by the  polarizability parameter $\kappa$, which accounts for deviations of the actual {\sc GDR} effects
% (circles in Fig.~\ref{fig:sigmasym})
to those predicted by the hydrodynamic model~\cite{migdal1944quadrupole,levinger} and is directly proportional
to the polarization potential in modern Coulomb-excitation analysis codes~\cite{gosia,tomdrake}.
% Empirically,  $\kappa$ values for  ground states,
% $\kappa(g.s.)$, can  be determined from $\sigma_{_{-2}}$ values~\cite{orce_review},
% \begin{equation}
% \sigma_{_{-2}}=
% % 2.38\kappa\times 10^{-4} A^{5/3} ~\mbox{fm$^2$/MeV} =
% 2.38\kappa ~A^{5/3} \mu\mbox{b/MeV},
% \label{eq:2p4}
% \end{equation}
% where $\sigma_{_{-2}}$ is the $(-2)$ moment of the total electric-dipole
% photo-absorption cross section, $\sigma_{_{total}}(E_{_{\gamma}})$~\cite{levinger2,migdal3},
% \begin{equation}
% \sigma_{_{-2}}=\int_{E_{\gamma_{min}}}^{E_{\gamma_{max}}} \frac{\sigma_{_{total}}(E_{_{\gamma}})}{E_{_{\gamma}}^{^2}}~dE_{_{\gamma}}.
% \label{eq:sigma-2int2}
% \end{equation}
% % determined from the 1988 photo-neutron cross-section evaluation with monoenergetic photons~\cite{atlas}.
% % which increases with the diffuseness of the nucleus.
% Typically,  $\sigma_{_{total}}(E_{_{\gamma}})$  includes
% $\sigma$($\gamma$,n), $\sigma$($\gamma$,2n) and $\sigma$($\gamma,np)$ photo-neutron~\cite{atlas} and $\sigma$($\gamma,p)$ photo-proton contributions~\cite{levinger2}
% % and Eq.~\ref{eq:sigma-2int2} should be integrated between 0 and $\infty$; although in practice, $E_{\gamma_{min}}$
% and $E_{\gamma_{max}}$ are,  respectively, the empirically available lower and upper endpoint $\gamma$-ray energies.
% (typically the neutron separation energy~\cite{atlas}) 
% The polarization potential $V_{pol}(t)$ influences the excitation cross section and may affect the determination of
Collective properties may be enhanced for $\kappa>1$,  whereas shell effects can be inferred from $\kappa<1$~\cite{cebo,cebo2,cebo3}.
% has no effect on collectivity but
% provides an indication of
% or a reduced oscillator strength~\cite{orce2023global,orce2023electric}.
% Accordingly, the value of $\kappa$ multiplies the polarization potential in modern Coulomb-excitation analysis codes~\cite{gosia,tomdrake}.
% This spurious enhancement of collectivity can be corrected from the quadrupole interaction
% $V_{_0}(t)$ by employing  an effective quadrupole interaction~\cite{hausser2},
% $V_{_{eff}}(t) =  ~V_{_0}(t)~ \left(~1-V_{pol}(t)\right)$,
% % which is
% used in modern Coulomb-excitation analysis codes~\cite{gosia,tomdrake}
% % such as {\sc GOSIA}~\cite{gosia}.
% % and  Winther$-$de Boer~\cite{tomdrake}.
% % to determine matrix elements.
% by simply modifying $\kappa$ in Eq.~\ref{ref:finalpol} for the case of interest.
Negligible polarizability effects were assumed in previous {\sc RE} measurements of $^{20}$Ne~\cite{Nakai,Schwalm,Olsen}.
% so-called $E1$ polarization 
% parameter of 0.0038 
% which arises from 
% Eq.~\ref{eq:2p4} 
% % and ought to be the default value in Coulomb-excitation codes 
% for $\kappa=1$~\cite{orce1}.
% in Coulomb-excitation codes.

In order to assess the polarizability of the ground state, $\kappa(g.s.)$,  in $^{20}$Ne
values of $\sigma$($\gamma,p)=165(25)$ MeV$\cdot$mb and $\sigma$($\gamma,n)=115(20)$  MeV$\cdot$mb~\cite{gorbunov} were
extracted in agreement with
% and $160(80)$ MeV$\cdot$mb~\cite{komar} have been observed,
% which are consistent with
1) the Thomas-Reiche-Kuhn ({\sc TRK}) sum rule, $\sigma_{_{total}}(E_{_{\gamma}})=60~NZ/A=300$ MeV~\cite{levinger2,lectures},
and 2) proton emission being the predominant decay mode in $A=4n$  self-conjugate nuclei~\cite{orce2022competition}.
A similar value of  $\sigma$($\gamma,p)=160(80)$ MeV$\cdot$mb has been observed~\cite{komar}.
The  proton-neutron yield ratio, $r_{pn}= \sigma(\gamma,p)/\sigma(\gamma,n)\approx165(25)/115(20)=1.4(3)$,
% where the uncertainties are in accord with literature values~\cite{gn},
% where a 10\% uncertainty for $\sigma(\gamma,n)$ is in agreement with literature values~\cite{gn}.
is in  agreement with predictions based on the evaporation model of Blatt and Weisskopf~\cite{blatt2012theoretical}, $r_{pn}=1.3$~\cite{orce2022competition}.
% calculated on the basis of compound nucleus formation, which increases exponentially as a function of $A$ for self-conjugate nuclei,
% giving $r_{pn}\approx2.6$ for $^{20}$Ne.
% , in  agreement with the estimated $\sigma(\gamma,p)/\sigma(\gamma,n)\approx165(25)/60(6)=2.8(5)$,
% where a 10\% uncertainty for $\sigma(\gamma,n)$ is in agreement with literature values~\cite{gn}.
The resulting $\sigma_{_{-2}}= 600(90)$ $\mu$b/MeV~\cite{gorbunov} ---  which also includes ($\gamma,\alpha)$ contributions ($\approx10\%$) ---
yields $\kappa(g.s.)=1.7(3)$.

\begin{figure*}[]
\begin{center}
\includegraphics[width=5.5cm,height=5cm,angle=-0]{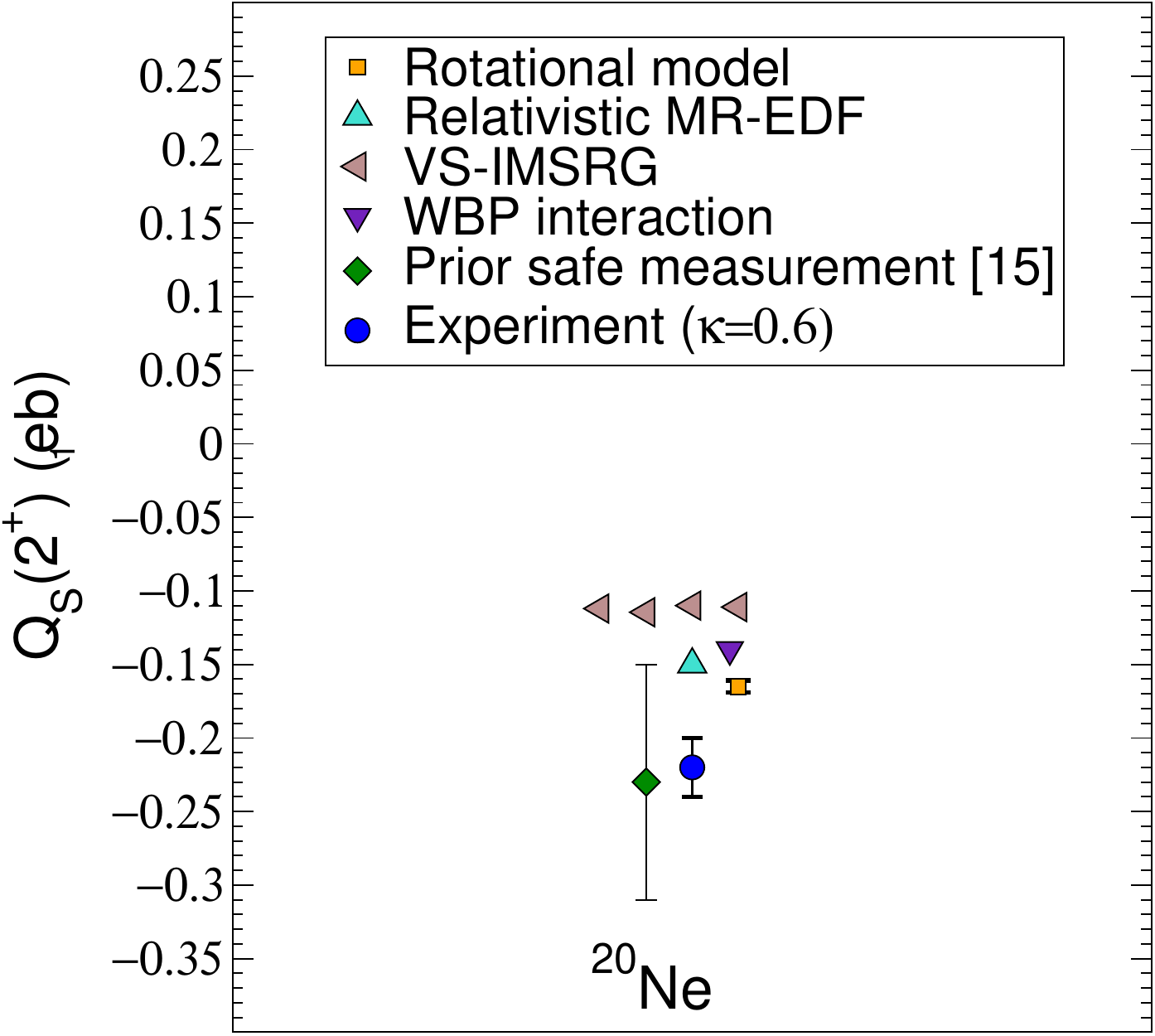}
\hspace{1.cm}
\includegraphics[width=7cm,height=5cm,angle=-0]{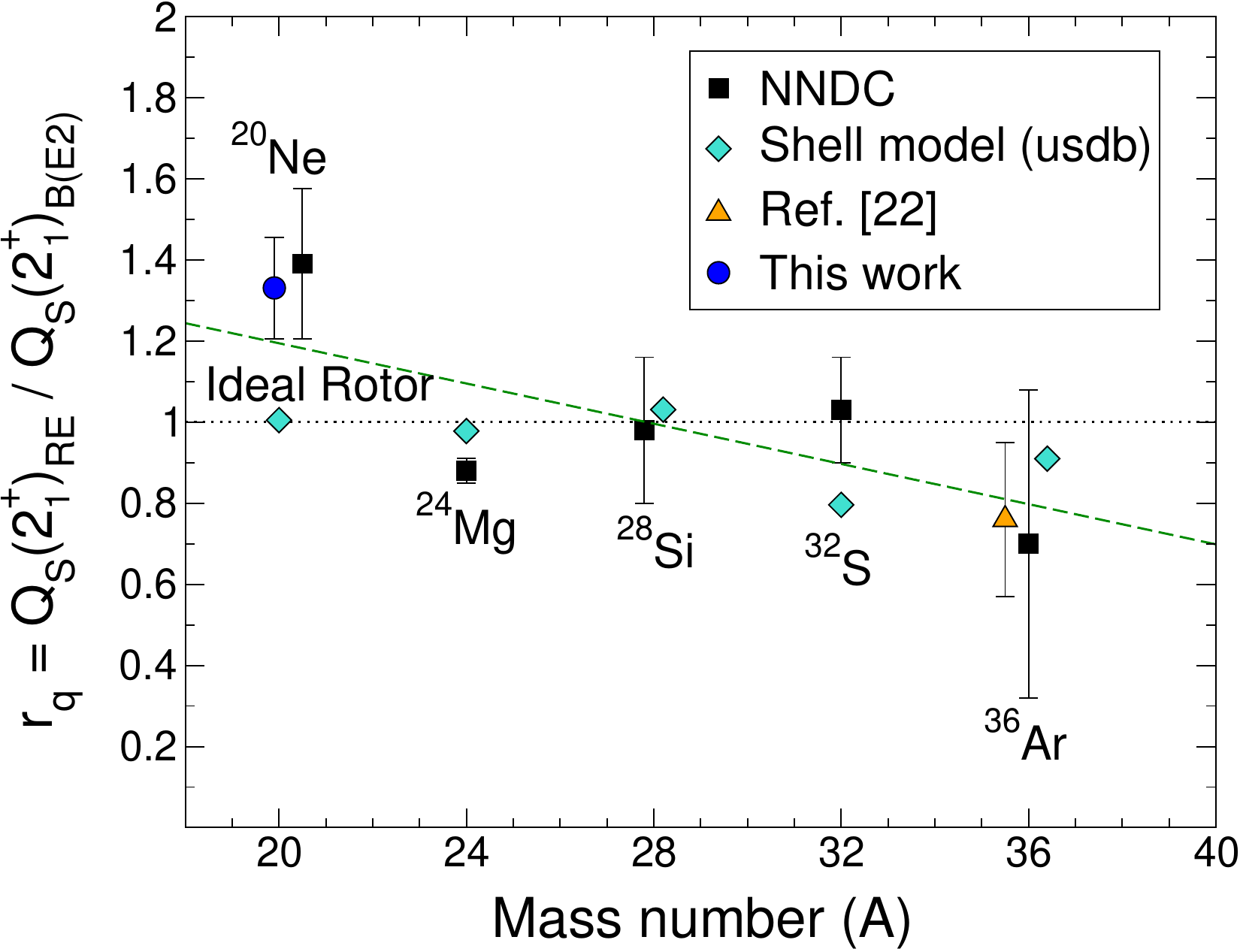}
\caption{The left panel shows  experimental (rotational model, {\sc NNDC}~\cite{Spear}, previous work at safe energies~\cite{Schwalm} and current work for $\kappa=0.6$) and theoretical  (relativistic MR-EDF, VS-IMSRG and WPB) $Q_{_S}(2^+_1)$ values in $^{20}$Ne   determined in the current work. For the VS-IMSRG calculations, the interactions are from left to right: PWA,
N4LO, N2LO$_{sat}$ and 1.8/2.0(EM).
% The  $Q_{_S}(2^+_1)$ values extracted from the {\sc NNDC}~\cite{nndc} and the rotational model are shown for comparison.
% The right panel shows the characteristic  $2D$ intrinsic nucleon density of the 2$_1^+$ state in $^{20}$Ne obtained with the MR-EDF model.}
% The excitation energy of the
% \emph{\sc GDR} built on excited states, $E_{_{GDR}}$, is given by
% % the combined power relation,
% $31.2A^{-1/3}+20.6A^{-1/6}$ MeV~\cite{GDRenergy}.}
The right panel shows experimental and theoretical $r_q$ values for self-conjugate nuclei within the $sd$-shell. For comparison,
a linear regression to the most accurate data is shown by a dashed line.}
\label{fig:theoryvsexp3}
\end{center}
\end{figure*}

Shell-model ({\sc SM}) calculations have additionally been performed in order to estimate $\kappa(g.s.)$
by combining the hydrodynamic model and the  $\sigma_{_{-2}}$ value calculated  from  second-order perturbation theory~\cite{orce2023global},
\begin{eqnarray}
\sigma_{{-2}}=\frac{16\pi^3e^2}{9\hslash c}\frac{1}{2J_{_{i}}+1} \sum_n \frac{\big| \langle i\parallel\hat{E1}\parallel n\rangle \big|^2}{E_{{\gamma}}},
% \approx 10~\sigma_{{-2}}, 
\label{eq:polar}
\end{eqnarray}
using the {\sc OXBASH} {\sc SM} code~\cite{oxbash} with the WBP interaction~\cite{warburton1992effective} (the original $usd$) and the $spsdpf$ model space.
A value of  $\kappa(g.s.)=1.7$ is calculated in agreement with the experimental value.
% by including $E1$ matrix elements from all $1\hslash\omega$ transitions
% connecting 212 1$^-$ states up to 30 MeV --- i.e. well above the {\sc GDR} region ---
% and 1520 1$^-$ bound states up to 76.7 MeV, values of $\kappa(g.s.)=2$ and 2.2 are predicted. 
Details of the calculations are presented in Refs.~\cite{orce2023global,orce2023electric}.
Assuming the same $\kappa$ value for the 2$^+_1$ state, $\kappa(2^+_1)=1.7$, the $\chi^2$  minimization yields  $\langle 2^+_1\mid\mid \hat{E2} \mid\mid 2^+_1\rangle=-0.28(3)$ eb and $Q_{_S}(2^{+}_{1})=-0.21(2)$ eb, as shown in the left panel of Fig.~\ref{fig:chqsurface}.

Nonetheless, growing evidence suggests that $\kappa$ may change from the ground  to excited states~\cite{hausser2,barker,orce_10Be,12C_kumar,orce2023electric}.
% Similar {\sc SM} calculations have been performed to determine $\kappa$ for the 2$^+_1$ state, $\kappa(2^+_1)$~\cite{hausser2,barker,orce_10Be,12C_kumar}.
% , where
% \begin{eqnarray}
% \kappa=\frac{S(E1)}{S_{_0}},
% \end{eqnarray}
% with $S(E1)$ defined as
% \begin{equation}
% S(E1) := \frac{1}{2J_i+1}~\sum_{n} W(11J_iJ_f, 2J_n) ~\frac{\langle i\mid\mid
% \hat{E1} \Vert n\rangle \langle n\Vert \hat{E1} \mid\mid
% f\rangle}{E_n - E_i}, \label{eq:se1}
% \end{equation}
% and the sum extends over all intermediate {GDR} states $|n\rangle$
% connecting the initial ground $\ket{i}$ and final excited
% $|f\rangle$  states with $E1$ transitions, and $W(11J_iJ_f, 2J_n)$
% is the Racah W-coefficient~\cite{racah} ($J_i=0$, $J_n=1$ and
% $J_f=2$ for this particular case).
% % It is noteworthy that the product
% % of two $\hat{E1}$ operators yields an $\hat{E2}$ operator; hence,
% % some of the isoscalar giant quadrupole resonance strength may appear
% % in the sum given in Eq.~\ref{eq:se1}.
% The result from Eq.~\ref{eq:se1} has to be normalized to the  units typically used
% for the hydrodynamic model~\cite{hausser2}{;} hence, $S_{_0}$ is defined as
% \begin{eqnarray}
% S_{_0}:=\sigma_{_{-2}} \eta_{_0} e^2= 0.00039\frac{A}{Z}\langle i\Vert
% \hat{E2} \Vert f\rangle e^2,
% \end{eqnarray}
% where $\sigma_{_{-2}}$ is given by Eq.~\ref{eq:2p4} and the
% dimensionless parameter $\eta_{_0}$ by~\cite{hausser2},
% \begin{eqnarray}
% \eta_{_0}=\frac{4\sqrt{\pi}}{3} \frac{\langle i\Vert \hat{E2} \Vert
% f\rangle }{ZeR^2}.
% \end{eqnarray}
% Both $S(E1)$ and $S_{_0}$ are given in units of $e^2$fm$^2$/MeV.
In fact, recent {\sc SM} calculations yield a smaller value of $\kappa(2^+_1)=0.6$ in $^{20}$Ne~\cite{orce2023electric}, which is the adopted value in this work.
% The  20\% uncertainty assigned for $\kappa(g.s.)=1.7$ from the difference between the theoretical and experimental values
% may be larger here.
Assuming $\kappa(2^+_1)=0.6$, the  $\chi^2$  minimization shown in the middle panel of Fig.~\ref{fig:chqsurface} yields similar values of
$\langle 2^+_1\mid\mid \hat{E2} \mid\mid 2^+_1\rangle=-0.29(3)$ eb and $Q_{_S}(2^{+}_{1})=-0.22(2)$ eb,
% % , as shown in Fig.~\ref{coulex} (yellow band).
% Finally, adding in quadrature the previous error for the diagonal matrix element with the error of $\kappa$ in the interval [2.0,3.0],
% a value of
% $\langle 2^+_1\mid\mid \hat{E2} \mid\mid 2^+_1\rangle=-0.24(4)$ eb and 
% $Q_{_S}(2^{+}_{1})=-0.19(^{+0.02}_{-0.03})$ eb together with
which includes a 3$\%$ systematic error due to quantal effects.
The right panel of Fig.~\ref{fig:chqsurface} shows
how $\langle 2^+_1\mid\mid \hat{E2} \mid\mid 2^+_1\rangle$ shifts towards less prolate deformed shapes with increasing $\kappa$ values.
% Table \ref{tab:2} shows the transitional and diagonal matrix elements together with the corresponding $Q_{_S}(2^{+}_{1})$ and $B(E2)\uparrow$
% values for $\kappa(2^+_1)=0.6$ obtained in this work.
% \begin{center}
% \begin{table}[!ht]
% \caption[$\langle 2^+_1\mid\mid \hat{E2} \mid\mid 2^+_1\rangle$, $\langle 0^+_1\mid\mid \hat{E2} \mid\mid 2^+_1\rangle$, $Q_{_S}(2^{+}_{1})$ and $B(E2)\uparrow$ for $\kappa(2^+_1)=0.6$]{$\langle 2^+_1\mid\mid \hat{E2} \mid\mid 2^+_1\rangle$, $\langle 0^+_1\mid\mid \hat{E2} \mid\mid 2^+_1\rangle$, $Q_{_S}(2^{+}_{1})$ and $B(E2)\uparrow$ for $\kappa(2^+_1)=0.6$ obtained in this work.}\label{tab:2}
% \centering
% \begin{tabular}{c c c}
% \hline\hline %inserts double horizontal lines
% $\kappa$ & $\langle 0^+_1\mid\mid \hat{E2} \mid\mid 2^+_1\rangle$ (eb) & $\langle 2^+_1\mid\mid \hat{E2} \mid\mid 2^+_1\rangle$ (eb)\\
% \midrule
% % inserts single horizontal line
% 0.6 & -0.29($^{+0.03}_{-0.03}$)  & 0.182($^{+0.005}_{-0.005}$)\\
% \hline\hline
% $\kappa$ & $ Q_{s}(2^{+}_{1})$ (eb) & $B(E2)\uparrow$ ($e^{2}b^{2}$)\\
% \midrule
% % inserts single horizontal line
% 0.6 & -0.22($^{+0.02}_{-0.02})$ & 0.0332($^{+0.0003}_{-0.0002}$)\\
% \midrule
% \end{tabular}
% \end{table}
% \end{center}
% , which corresponds  to a weighted average of $Q_{_S}(2^+_1)=-0.21(2)$ eb.

This precise experimental result can now be used to benchmark nuclear models and modern state-of-the-art theoretical calculations.
As shown in the left panel of Fig.~\ref{fig:theoryvsexp3}, this new result (circle for $Q_{_S}(2^{+}_{1})=-0.22(2)$ eb) presents a
% now in agreement with
% some previous theoretical calculations
% which cluster around  $\approx-0.15$ eb,
% and
2.735$\sigma$ standard deviation with respect to the one extracted with the rotational model,
$Q_{_S}(2^+_1)_{B(E2)} = \pm 0.165(4)$ eb (square), which is arguably one of the largest deviations observed in nuclei.
In addition, a large $r_q=1.33(12)$ value is determined, which  is clearly inconsistent with that expected for axially-symmetric rotors  ($r_q=1$).
The right panel of Fig.~\ref{fig:theoryvsexp3} shows the $r_q$ values for self-conjugate nuclei in the $sd$ shell.
% Shell-model (SM) calculations using different 2N and 2N+3N chiral forces are shown on the left panel of Fig.~\ref{fig:theoryvsexp3} 
% (left triangles), which present consistent results, $Q_{_S}(2^+_1)\approx-0.11$ eb, although underestimate the experimental value. 
In the {\sc SM} calculations the $r_q$ ratio does not depend on the effective charges. Values of $r_q\approx1$ (diamonds) consistent with ideal rotors are computed with the {\sc WBP}~\cite{warburton1992effective} and the $usdb$~\cite{sdeff} interactions. More details on these {\sc SM} calculations can be found in Ref.~\cite{orce2021reorientation}.
Nevertheless, large deviations from axial symmetry are evident at both extremes of the $sd$ shell,
suggesting a missing factor.
% , which requires additional investigation.

\begin{figure}[]
\begin{center}
\includegraphics[width=6cm,height=4.7cm,angle=-0]{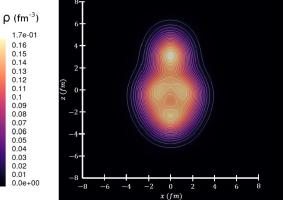}
\caption{Characteristic  $2D$ intrinsic nucleon density of the 2$_1^+$ state in $^{20}$Ne obtained with the MR-EDF model.}
% The excitation energy of the
% \emph{\sc GDR} built on excited states, $E_{_{GDR}}$, is given by
% % the combined power relation,
% $31.2A^{-1/3}+20.6A^{-1/6}$ MeV~\cite{GDRenergy}.}
\label{fig:cluster}
\end{center}
\end{figure}

We have further performed {\it ab initio} calculations of energies and quadrupole moments in $^{20}$Ne with the valence-space formulation of the in-medium similarity renormalization group (VS-IMSRG)~\cite{hergert,stroberg,tsukiyama,bogner,stroberg2}. In this approach, we construct a valence-space Hamiltonian, based on two- (NN) and three-nucleon (3N) forces from chiral effective field theory. In particular we use  the ``EM1.8/2.0'' and ``PWA'' NN+3N interactions~\cite{hebeler,simonis,simonis2,morris,jason} ---
the former of which reproduces ground-state and separation energies up to the tin region from constraints to only few-body data~\cite{morris,jason} ---  as well as N$_2$LO$_\mathrm{sat}$~\cite{ekstrom} and N$_4$LO local-non-local interaction of Ref.~\cite{Leistenschneider}, which well reproduces radii to the nickel region~\cite{ekstrom,groote}.
We use the Magnus procedure~\cite{hergert,morris2} to decouple an $sd-$shell Hamiltonian as well as the $^{16}$O core energy. The effects of 3N forces between the $^{20}$Ne valence nucleons are included via the ensemble normal ordering technique~\cite{stroberg2}. Finally, we use the approximate unitary transformation to decouple a consistent $E2$ valence-space operator~\cite{parz}, where no effective charges are used or needed, in principle.  Similar results of $Q_{_S}(2^+_1)\approx-0.12$ eb are computed, as shown in the left panel of Fig.~\ref{fig:theoryvsexp3} (left triangles).
This underestimation is not surprising as ab initio $E2$ transition rates are smaller than experimental ones, regardless of the initial interaction.
This is likely due to the complex collective structures which require many particle-hole excitations to capture sufficiently and are a particular challenge for most methods~\cite{garcia,jack,klose,heil}. 
% However, trends are  in capturing the highly collective physics of this transition.
% the trends typically agree well with experiment [19]. 
Recent {\it ab initio} calculations in the neon isotopes with the projected generator coordinate method ({\sc PGCM}) at a harmonic-oscillator basis size of $N_{max}=4$
compute a  larger value of $Q_{_S}(2^+_1)=-0.183$ eb in $^{20}$Ne~\cite{sarma2023ab}. Further studies of the origin and solution to this missing $E2$ strength are underway~\cite{jack2}.
A slightly smaller value of $Q_{_S}(2^+_1)=-0.143$ eb (down triangle in Fig.~\ref{fig:theoryvsexp3}) is determined using further {\sc SM} calculations with the {\sc WBP} interaction
and average effective charges in the $sd$-shell~\cite{sdeff}. Similar results are obtained with the $usdb$ and $usdc$ interactions~\cite{usdc}.
An ingredient that is not explicitly included in these {\it ab initio} and phenomenological interactions is  $\alpha$ clustering.

Furthermore, we have performed multi-reference energy density functional (MR-EDF) calculations \cite{bender2003,niksic2011,robledo2019,edftextbook} based on the relativistic Hartree-Bogoliubov (RHB) model \cite{vretenar2005,meng2006}, using DD-PC1 functional \cite{nikvsic2008relativistic} and separable pairing force \cite{tian2009} as global effective interactions. A total of $118$ RHB states with a wide range of quadrupole and octupole deformations were first projected onto good values of angular momenta, particle numbers, and parity, and further mixed within the generator coordinate method framework~\cite{hwg, marevic3}. The calculated $2_1^+$ state at $1.84$ MeV
exhibits a predominantly prolate $^{16}$O+$\alpha$ structure --- the bowling pin shape shown in Fig.~\ref{fig:cluster} --- and decays to the ground state with a quadrupole transition strength of $B(E2; 0_1^+ \rightarrow 2_1^+) = 0.029$ $e^2b^2$, in agreement with the experimental value.
The corresponding $Q_{_S}(2^+_1)=-0.15$ eb (triangle up on the left panel of Fig.~\ref{fig:theoryvsexp3}) and $r_q=0.97$ values 
are still too small compared with the experimental values determined in this work.
% ($\kappa=0.6$).
% The rise of such a pronounced prolate shape could be the reason behind the large $r_q$ value.
% determined experimentally for $^{20}$Ne.  
% The fact that the $Q_{_S}(2^+_1)_{B(E2)}$ value extracted 
% from the rotor model and $Q_{_S}(2^+_1)$ are now consistent with each other (within $0.5\sigma$) also supports this point. 
% Nonetheless, the $^{16}$O+$\alpha$ cluster could not exist and the alpha particle could therefore not induce a large prolate shape if the octupole degree of freedom was not prominent (i.e., breaking of the parity symmetry), which is probably the reason why these nuclei 
% do not present such large $E2$ strengths with respect to nuclides in the rare-earth or $A\sim180$ mass regions. 
Ongoing efforts to quantify the uncertainties of EDF-based models \cite{edftextbook,dobaczewski2014,mcdonnell2015}, particularly when extended to the multi-reference level, may enable a more
rigorous test.
% the predictability of global EDFs
% against the ever-growing bulk of experimental data.
% It is interesting to note that previous work are also in agreement with our results~\cite{Spear,20Ne_gogny}. 
% For instance, the Angular Momentum Projected Generator Coordinate Method (AMP-GCM) -- 
% with the quadrupole moment as collective coordinate and the Gogny force (D1S)~\cite{20Ne_gogny} -- calculates $Q_{_S}(2^+_1)=-0.1675$ eb (right triangle in Fig.~\ref{fig:theoryvsexp3}).  

% In fact, a slightly larger $\kappa$ value may be expected from additional photo-proton contributions (80 MeV$\cdot$mb still missing to 
% exhaust the TRK dipole sum), 
% % ~\cite{gorbunov,komar}
% which could shift the $Q_{_S}(2^+_1)$ value toward an even less-deformed prolate shape. 

% The low-lying structure of $^{20}$Ne is marked by the admixture of  shell-model and cluster-like configurations~\cite{Hiura72,Fujiwara80,Itagaki11}, with pronounced nucleon localization being present already in its ground state~\cite{Kanada,Ebran,The low-lying structure of $^{20}$Ne is marked by the admixture of  shell-model and cluster-like configurations~\cite{Hiura72,Fujiwara80,Itagaki11}, with pronounced nucleon localization being present already in its ground state~\cite{Kanada,Ebran,Zhou12}. }.

% in Fig.~\ref{coulex} 
% yielding  better agreement 
% between {\sc RE} measurements, theory and rotational model.

% Consequently, 
% this work shows that the nuclear polarizability plays an important role  when extracting 
% nuclear collective properties,
In conclusion, we present the most sensitive and precise reorientation-effect measurement of $Q_{_S}(2^+_1)=-0.22(2)$ eb in $^{20}$Ne up to date
determined at safe energies with a heavy target and backward scattering angles through the $\chi^2$ minimization of the integrated yields.
% using the Coulomb-excitation least-squares search code {\sc GOSIA}.
We further quantify the competing second-order contribution arising from the electric dipole polarizability through 1$\hbar\omega$ shell-model calculations.
% The latter has never been measured for excited states of even-even nuclei, although this could, in principle, be done with dedicated experiments using modern spectrometers such
% as the {\sc GAMKA} spectrometer in South Africa~\cite{GAMKA} and a broad range of beam energies and scattering angles~\cite{eichler}.
Large deviations from the rotor-model and various modern nuclear theory approaches are found, including {\it ab initio} and {\sc MR-EDF} calculations done in this and prior work.
The presence of a bowling pin shape for the 2$^+_1$ state in $^{20}$Ne calculated with {\sc MR-EDF} suggests that $\alpha$ clusters should be explicitly included
in modern {\it ab initio} calculations~\cite{frosini2022multi}.
In fact, a recent calculation by D. Lee reproduces the quadrupole moment presented here when considering $\alpha$ clusters in the interaction~\cite{dlee}. \\

We acknowledge the accelerator group at iThemba LABS for masterful delivery of pure $^{20}$Ne beams.
CVM and JNO thank
L. Gaffney,  N. Warr, D. Lee, L. Robledo and the late D. Schwalm for interesting physics discussions and providing relevant information.
This work was partly performed under the auspices of the U.S. Department of Energy by the Lawrence Livermore National Laboratory
(LLNL) under contract No. DOE-AC52-07NA27344, the National Science Foundation (NSF) under grant PHY-1811855,
% and the National Research Foundation (NRF) for financial support. 
and the South African National Research Foundation (NRF) under Grant 93500
% the SANHARP and
% % the MANuS/MatSci Honours/Masters programmes, and
% the SA-CERN Collaboration.

\bibliographystyle{apsrev4-2}% Produces the bibliography via BibTeX.

\bibliography{20neref}% common bib file

\end{document}